\documentclass[sigplan,9pt,screen,natbib,nonacm]{acmart}
\setcopyright{none}
\pdfoutput=1

\usepackage{hyphenat}

\usepackage{mdwtab}
\usepackage{mdwlist}
\usepackage{mathenv}
\usepackage{amsmath}
\usepackage{comment}
\usepackage{url}

\usepackage{listings}

\lstdefinestyle{sOcaml}{language=[Objective]Caml,
  literate={+}{{$+$}}1 {/}{{$/$}}1 
           {=}{{$=$}}1
           {>}{{$>$}}1 {<}{{$<$}}1
           {<>}{$\not=$}1
           {>>}{$\gg$}1
           {->}{{$\rightarrow$}}2 {>=}{{$\geq$}}2 {<-}{{$\leftarrow$}}2
           {<=}{{$\leq$}}2
           {==>}{{$\mapsto$}}2
           {|}{{$\mid$}}1
           {|>}{$|\!{>}$}1
           {'a}{$\alpha$}1
           {'b}{$\beta$}1
           {'c}{$\gamma$}1
           {'d}{$\delta$}1
           {'w}{$\omega$}1
           {'r}{$\rho$}1
           {'state}{$\sigma$}1
           {:=}{\ensuremath{\mathrel{{:}{=}}}}2
           {...}{\ldots}2
           {\#\#+}{\color{red}}1
           {\#\#-}{\color{black}}1
           {\#\#\#}{{$\leadsto$}}3
}
\lstnewenvironment{code}[1][]{\lstset{#1,basicstyle={\sffamily\small}}}{}

\lstMakeShortInline[columns=fullflexible]|

\newcommand{\aside}[1]{\ignorespaces}

\begin{document}
\title{Free Variable as Effect, in Practice}

\author{Oleg Kiselyov}
\orcid{0000-0002-2570-2186}
\affiliation{%
  \institution{Tohoku University}
  \country{Japan}}
\email{oleg@okmij.org}

\begin{abstract}
Variable environment is the time-honored way of making sense of free
variables, used in programming language theory as well when writing 
interpreters and some compilers. Algebraic effects give another way,
as was pointed already at HOPE 2017. Although a
theoretical curiosity, it may have surprising practical benefits: 
a new way of writing compilers, with the incremental type-checking,
with easy variable usage, leaf
function analyses. This work-in-progress report
prototypes and illustrates the idea. 
\aside{It also touches on a new way of
thinking about functions.}
\end{abstract}
\maketitle

\section{Introduction}
\label{s:intro}

Whenever one writes an interpreter or a compiler, or studies logic,
model theory, programming language theory~-- one soon has to face
variables. The well-known way to deal with them is by introducing
an \emph{environment}, or variable assignment in logic. There is,
however, another, more general approach, related to algebraic
effects. One should not be too surprised. Algebraic effects originate
from studying terms with variables, and equations on them: free
algebras.

The algebraic effect approach was already described at HOPE 2017
\cite{having-effect} and elaborated at \cite{free-vars-effect}. It was
explored and scaled up in the study of rigorous,
realistic and interesting reasoning with effects \cite{NBEA}.

Here we look at an unexpected practical side, in interpreting or
compiling languages. It has occurred to me as I was teaching compiler
class developing the complete compiler to x68-64 assembly
feature-by-feature, and in tagless-final style.

The first benefit is the ability to evaluate intermediary expressions
and report errors soon, before the whole program is parsed~-- hence
reducing the amount of memory for intermediary data and improving
latency. The approach also facilities variable usage and leaf function
analyses, indispensable in compilation.

\aside{Returning to theory, We also look at a new meaning of
  functions.}

\section{Interpreting Languages with Variables}
\label{s:interpreter}

To explain the idea, let's write an interpreter, which we later turn
into a compiler by changing the domain of
interpretation.\footnote{\relax
The complete code accompanying the paper is available at
 \url{https://okmij.org/ftp/Computation/var-effect/}}
In rough strokes the development, however simplified,
actually follows the compiler
class.

We start with the simplest interpreter: the source language has
only integers and addition. The language should hopefully be clear from
  the grammar, in the (ocaml)yacc form. We borrowed this example
from the ocamllex/ocamlyacc chapter of the OCaml Reference 
\cite[\S15.6]{OCaml}.
\begin{code}[mathescape=false]
exp:   INT                      { int $1 }
       | exp PLUS exp      { add $1 $3 }
       | LPAREN exp RPAREN { $2 } ;
\end{code}
The grammar defines the concrete syntax of the language.  The semantic
actions |int| and |add| are arranged in a separate module with the
following signature, which in effect, defines the abstract syntax.
\begin{code}
module type LangInt = sig
  type repr                               (* representation type *)

  val int : int -> repr
  val add : repr -> repr -> repr

  type obs                                (* observation type *)
  val observe : repr -> obs
end
\end{code}
Where |repr| is the domain of the interpretation.

Here is one implementation of the signature.
\begin{code}
module EvalInt = struct
 type dom = int
 type repr = dom

 let int x = x
 let add x y = let s = x + y in printf "   => %d\n%!" s; s

 type obs = unit
 let observe x = string_of_int x |> print_endline
end
\end{code}
The value domain is |int| (OCaml integers), which is also the domain
of interpretation |repr|. The function |observe| is invoked after the
parsing is finished; it observes the |repr| value representing the
result of the whole program, by printing it. We also made the
interpreter to print the (intermediate) results, of each addition
expression. As we shall see, it is a good diagnostic for the
evaluation order.\footnote{as well as memory requirements: deferring a
  computation needs memory to store what is to compute later.}

Let's add variables. We add the productions
\begin{code}[mathescape=false]
 | IDENT                                 { var $1 }
 | LET IDENT EQ exp IN exp { let_ ($2,$4) $6 } 
\end{code}
to the parser, and likewise extend the abstract syntax. By `extending'
we mean creating a new version re-using the old code~-- \emph{in its already
compiled form, and without any copy-pasting or editing}.

\begin{code}
module type LangLet = sig
  include LangInt

  type name = string

  val var : name -> repr
  val let_ : name * repr -> repr -> repr
end
\end{code}
Its implementation also re-uses |EvalInt|, but re-defines all
operations:\footnote{Here
$\gg$ is left-to-right function composition}
\begin{code}
module EvalEnv = struct
 type dom = EvalInt.dom
 type name = string
 type env = (name * dom) list
 type repr = env -> dom

 let ans   : dom -> repr = fun v -> fun _env -> v
 let lift2 : (dom->dom->dom) -> (repr->repr->repr) = fun op e1 e2 ->
    fun env -> op (e1 env) (e2 env)

 let int = EvalInt.int >> ans
 let add = lift2 EvalInt.add
 let var : name -> repr = List.assoc
 let let_ : name * repr -> repr -> repr = fun (n,b) body -> 
   fun env -> body ((n,b env) :: env)

 type obs = unit
 let init_env : env = []
 let observe x = x init_env |> EvalInt.observe
end
\end{code}
To handle variables, we introduce the variable environment |env|~-- here, 
the associated list
of variable names and their meanings~-- as explained in every textbook
about interpreters. The domain of interpretation is now a function
from |env| to the value domain, to which the earlier semantic
functions are lifted. Again, the interpretation is
completely standard and explained in every textbook on this topic.

What the textbooks rarely point out is an undesirable change. In the
original |EvalInt|, the (intermediate) result of a sub-expression is
printed as soon as it is parsed. When we enter |"1+2\n+3\n+4\n"| we see
the partial sums printed as soon as we hit `Enter'. |EvalInt|
indeed works as the familiar desk calculator.\footnote{In fact, it
  makes a simpler and clearer example than the one in the ocamlyacc's
  reference manual.}

|EvalEnv| is different: As we enter the same
|"1+2\n+3\n+4\n"|, nothing is printed. It is only when we terminate
the input and tell the parser the whole program is finished that we
see the results. Whereas |EvalInt| interprets \emph{as} the program
is being parsed, |EvalEnv| does the real work (summation) only
\emph{after} the whole program has been parsed. It is not hard to see
why.

The meaning of |1+2| in the |EvalInt| semantics is |3| (computed
compositionally). In the |EvalEnv| semantics, the same expression has the
meaning
\begin{code}
let m12_env : EvalEnv.repr = 
     fun env -> (fun _ -> 1) env + (fun _ -> 2) env
\end{code}
which is a function.
Its body is not evaluated until it receives the |env| argument~--
even if the argument is not needed. That argument, the initial
environment, is passed by |observe| only when the entire program is
parsed. 
One may notice that |m12_env| has the structure
of the corresponding source expression |1+2|~-- obviously, since
 the meaning assignment is a homomorphism. The meaning is a function (closure)
that references the meanings of |1| and |2|, which are also closures.
In effect, |m12_env| is a parse tree of the source expression |1+2|~--
in a form of closures and taking hence more memory compared to a data
structure. This parse tree is
interpreted upon the final observation.

What |EvalEnv| gained, however, is handling programs with variables like
|(1+2)+x|~-- which cannot be evaluated until we
receive the environment and look up the value of |x|.
Still, the sub-expression |(1+2)| could be interpreted on
the spot. How to make it happen?

\subsection{Variable as an Effect}
\label{s:var-eff}

When dealing with expressions like |(1+2)+x|, we need to know what
value corresponds to |x|. We can just ask. The meaning of an
expression is then either an answer |A(v)|, or a question |Q(n,k)|
about the value of the variable |n|, to be continued as |k|,
perhaps asking further questions until the final answer. We hence
introduce the following variable effect (which is the Free monad
implementation of the Reader effect,
and entirely standard):
\begin{code}
module VarEff = struct
 type name = string
 type 'd t = A of 'd | Q of name * ('d -> 'd t)

 let ans : 'd -> 'd t = fun v -> A v
 let var : name -> 'd t = fun n -> Q(n,ans)

 let rec lift2 : ('d->'d->'d) -> ('d t -> 'd t -> 'd t) = fun op e1 e2 ->
  match (e1,e2) with
  | (A v1, A v2) -> A (op v1 v2)
  | (Q (n,k), e2) -> Q (n, (fun v -> lift2 op (k v) e2))
  | (e1, Q (n,k)) -> Q (n, (fun v -> lift2 op e1 (k v)))

 let lift : ('d -> 'd t) -> ('d t -> 'd t) = ...

 let handle_var : ('d -> 'd t) -> (name -> 'd option) -> 'd t -> 'd t = ...
 let letv : (name * 'd) -> 'd t -> 'd t = fun (n,v) ->
  handle_var ans (function n' when n'=n -> Some v | _ -> None)
 let top_hand : 'd t -> 'd = function A v -> v
end
\end{code}
A binary operation on two expressions |lift2 op| checks to see if both
operands have the answer. If so, the operation |op| can be performed
right away. Otherwise, |lift2| propagates operand's questions.
Eventually, the questions have to be answered, which is the job of a
handler. The handler |handle_var| is the mapping/fold over the
denotation (|'d t| tree). Its particular instance |letv|
replies to questions only about the given name, propagating all others.
The domain of interpretation is now |dom VarEff.t|, to which the
semantic functions are lifted:
\begin{code}
module EvalEff = struct
 module V = VarEff
 type dom = EvalInt.dom
 type repr = dom V.t

 let int = EvalInt.int >> V.ans
 let add = V.lift2 EvalInt.add
 let var = V.var

 let let_ : name * repr -> repr -> repr  = fun (n,b) body ->
    V.lift (fun v -> V.letv (n,v) body) b

 type obs = unit
 let observe x = V.top_hand x |> EvalInt.observe
end
\end{code}
As expected, |let_| acts as a handler, answering questions about
its bound variable, and propagating all other questions up.
One may show,
using the technique in \cite{NBEA}, that |EvalEff.repr| has the same
equational theory as |EvalEnv.repr|~-- that is,
|EvalEff| is extensionally equivalent to |EvalEnv|. Still,
|"(1+2)\n+x\n"| and |"x+(1+2)\n+3"| now print the result of interpreting
|1+2| right away, without waiting for the whole program to be
parsed. Furthermore, when we enter the program
\begin{code}
let y = let x = 1 + 2 
      in x + x + 3 in 
y + 1;;
\end{code}
we not only see |1+2| being evaluated right away, but also
|x+x+3| being evaluated as soon as it has been parsed, at the end of
the second line.\footnote{One can see that for themselves by compiling and
  running the code in the directories \textsf{step2} and
  \textsf{step3} in the accompanying code. The former implements the
  environment and the latter effect semantics for variables.
}
Questions about local variables can therefore be answered quickly,
without waiting for the whole program be
parsed.\footnote{However, straightened-out 
let-expressions are right-associated. Therefore, their parsing finishes
only at the end of the program.}

|EvalEff| offers further opportunities for optimization: if the body
of a let-expression has |A v| as its interpretation (denotation)~--
that is, not a question~-- the
body has not needed the value of the bound variable. 
We have hence come upon an easy
way to determine the usage of bound variables, which is valuable in
compilation, as we shall see in the next
section.

The variable-as-effect approach scales up to functions~-- 
as was in effect shown already in
\cite{having-effect}. Here are two sample programs\footnote{see
  \textsf{step4} in the accompanying code.}
\begin{code}
let x = 1 in 
let fun f(y) = x + y in
let x = 2 in f(2)

let x = 1 in 
let fun f(y) = x + y in 
let fun g(x) = f(x) in 
g(2)
\end{code}
Since a variable dereference is an effect, to be handled by a
dynamically enclosed handler, one may wonder if we are really
implementing lexical rather than dynamic binding. As was shown already in 
\cite{having-effect} and elaborated in \cite{free-vars-effect}, 
variable-dereference-as-effect does support lexical binding, 
with some work. Generally, a mechanism to capture the current dynamic
environment is needed. The current implementation uses a simpler 
approach: handling the body of a function in the handling
environment of its definition rather than of its invocation. Therefore,
both sample programs evaluate to |3|.

\section{Compilation}
\label{s:compilation}

The ability of |EvalEff| to evaluate as soon as possible, without
waiting for the whole program to be parsed is especially valuable in
compilation, where it translates to reporting type and other errors
early and reducing memory footprint. There is another benefit, hinted
earlier: the ease of variable use analyses, which are needed for
memory/register allocation. This section demonstrates both benefits. 

First, we turn our interpreter into a compiler, to Wasm. We change the
interpretation domain
 from |int| to a sequence of Wasm instructions that
leave the |int| result on the stack. 
\begin{code}
module EvalInt_wasm = struct
  type dom = Wasm.instr
  type repr = dom

  let int = Wasm.I32.const
  let add x y = Wasm.(exp [x; y; I32.add])

  type obs = unit
  let observe x = 
    let open Wasm in
    wasm_module [func ~result:I32 [x]] |> observe
end
\end{code}
We rely on the module |Wasm|: tagless-final embedding of 
Wasm.\footnote{see the directory \textsf{wasm} in the
  accompanying code.}
The new |EvalInt_wasm| is quite like
|EvalInt|, structurally. 
It interprets |"1+2+3"| as:
\begin{code}
i32.const 1   i32.const 2   i32.add   i32.const 3   i32.add
\end{code}
Just as we lifted |EvalInt| to |EvalEff| in \S\ref{s:var-eff}, 
we lift |Eval_wasm|;
the result, to be called |Eval_var|, is |EvalEff| with |EvalInt|
replaced with |Eval_wasm|. One may now compile programs with local
variables; for example, 
\begin{code}
let x=10+11 in 1+x+x+3
\end{code}
produces:
\begin{code}
i32.const 1   i32.const 10   i32.const 11   i32.add   i32.add
i32.const 10   i32.const 11   i32.add   i32.add   i32.const 3   i32.add
\end{code}
The variable |x| turns out substituted with its bound
expression: the let-binding got inlined.
One should not be too surprised:
after all, variables are like named `holes'
in the domain, with let-expressions telling how to fill the holes.
Such behavior of let-expressions~-- effecting sharing in the
compiler rather than in the object code~-- is well-known in code
generation \cite{swadi-monadic}.

To properly compile let-expressions, allocating storage (Wasm locals)
for bound variables, we lift |Eval_var| one more time.\footnote{see 
\textsf{step7}, in particular, \textsf{eval.ml} in that directory.}
In other words,
we generate Wasm with `holes', to be filled with the names of the
allocated Wasm locals. The allocation is performed after a
let-expression is compiled and the variable usage in its body is
determined.  Strictly speaking, the compilation becomes two-pass. However,
the first pass generates as much Wasm code as possible. Local
let-expressions can even be compiled entirely before the end of
parsing of the whole program.

The let-handler is particularly notable:
\begin{code}
let letv : name * dom -> repr -> repr  = fun (n,v) b ->
  let cnt  = ref 0 in                   (* usage count of n *)
  let vars = ref [] in                  (* other variables used *)
  let lkup = function
    | n' when n = n' -> incr cnt; Some (V.var n)
    | n' -> if List.mem n' !vars then () else vars := n' :: !vars; None
  in
  let ret res =
   if !cnt = 0 then V.ans res  (* no need to allocate anything *)
   else if !cnt = 1 then V.ans (Eval_var.let_ (n,v) res) (* inline *)
   else
    (* request allocation, reporting n and the list of alive,
       hence conflicted variables *)
  in V.handle ret lkup b
\end{code}
As the handler answers questions about its bound variable, it counts
them. At the end, it knows how many times the bound variable has been
accessed. If zero, there is no need to allocate storage for the
variable. (If the source language has no side effects, as ours
currently, we may even skip compiling the bound expression).
If the variable was used only
once, we substitute it with the bound expression, using |Eval_var|'s
let-machinery to do the substitution. Again, no storage allocation is
needed.  The letv-handler also watches for other variable requests,
and learns of all free variables in its managed expression. Their list
is reported to the allocator: these are conflicts, i.e., their storage
must be disjoint. We thus obtain all the information (variable usage
and conflicts) needed for storage allocation; see the source code for
details.

For example, the program
\begin{code}
let x = 1 + 2 in let y = x + 1 in let z = y + x in z + z + y
\end{code}
compiles to the following Wasm module
\begin{code}[mathescape=false]
(module  (func  (export "start" )  (result i32 ) 
(local $t_1  i32) (local $t_2  i32) 
(i32.const 1) (i32.const 2) i32.add local.set $t_1 local.get $t_1
(i32.const 1) i32.add local.set $t_2 local.get $t_2 local.get $t_1 i32.add
local.set $t_1 local.get $t_1 local.get $t_1 i32.add local.get $t_2 i32.add))
\end{code}
The variables |x| and |z| share the same Wasm local |t_1|.

Let us add functions~-- for simplicity, second-class top-level 
functions whose bodies have no free variables aside from the
arguments (since functions are second class, their names are distinct
from ordinary variable names).\footnote{Compiling functions with
  `open bodies' is rather challenging: Wasm intentionally prohibits
  accessing locals from a different function. To use locals as much as
  possible we would need an extensive variable use analysis, which
  should be feasible in our approach. This is the topic for future work.}
Here is an example:
\begin{code}
let fun f(x) = x + 2 in
let fun g(x,y) = f(y) + x in
f(g(1,2))
\end{code}
Since functions may take several arguments, there
comes the possibility of applying a function to a wrong number of
arguments~-- which is a type error. We should report it at
the compilation time.

The language with top-level second-class functions |Lang2Fun|
is the extension of |LangLet| with function calls and function
declarations: 
\begin{code}
module type Lang2Fun = sig
 include LangLet

 val call  : name -> repr list -> repr

 type fundecl                   (* function declaration *)
 val defun : name * name list * repr -> fundecl

 type defns                      (* a sequence of fundecl *)
 val defn_empty : defns
 val defn_add : defns -> fundecl -> defns

 type topform
 val top_exp : defns -> repr -> topform
 val topf_observe : topform -> obs
end
\end{code}
Here, |defun| interprets a declaration (the function name, the list of
argument names and the function body) as |fundecl|. Since functions
may only be declared at top-level and may not refer to outside
variables, all function declarations have to appear 
at the beginning of the program, followed
by the top-level expression (main program body)~-- which is what
|topform| signifies. The compilation for function bindings and
function calls is not much different from what we have seen for
integer-type let-expressions. A question about a function name is
answered with
its type (i.e., arity) and the Wasm name\footnote{Function names may
  be re-defined but Wasm names are unique.} (needed to generate the
Wasm call instruction). We refer to the accompanying code for details
(see the directory \textsf{step8}).

We have claimed that the effect semantics for variable and function
names enables incremental type checking and the early reporting of
errors. Let us see.  First, consider the OCaml code:
\begin{code}
let f(x) = x + 2
let g(y) = f(y,1) + y
f(g(1XXX
\end{code}
with two problems. On line 2 the function
|f| is invoked with a wrong number of arguments. Then there is a
parse error on line 3. Although it occurs later in the code, it and
only it is reported by the OCaml compiler:
\begin{code}
3 | f(g(1XXX
           ^^^^
Error: Invalid literal 1XXX
\end{code}
Indeed, an OCaml program must first be completely parsed, and only
then type-checked. When writing or refactoring code, however,
one would have liked to type check fragments (definitions) as soon as they are
finished, before the whole program is completed.

In contrast, if we submit the similar code
\begin{code}
let fun f(x) = x + 2 in
let fun g(y) = f(y,1) + y in
f(g(1XXXX
\end{code}
to our compiler, we get the compilation error about the first problem:
\begin{code}[keywords=]
Function f requires 1 arguments but was invoked with 2
\end{code}
In fact, if we feed the code into the compiler line-by-line, we notice that the
error is reported right after the second line is entered~-- before the
third, ill-formed, line is even input.

\section{Conclusions}

In the environment semantics the meaning of an
expression is a function from the environment, which is opaque and
cannot be examined. We cannot tell which variables in the environment
have actually been used, and how many times. Algebraic effects make the
denotation more observable: a handler can watch questions and find out
which variables have been asked about, and how many times. Thus we
obtain the variable usage analysis in the ordinary course of
compilation, almost for free, so to speak.

It remains to be seen how this promise holds for a real compiler for
a realistic programming language. I intend to find it out by trying
this technique out in the new installment of the compiler class
(which is underway).

\begin{acks}
I thank Chung-Chieh Shan for helpful discussions, and the reviewers
and participants
of the HOPE 2023 workshop for many insightful comments.
\end{acks}

\bibliographystyle{plainnat}
\bibliography{exteff.bib}
\end{document}